\documentclass[conference]{IEEEtran} 
\usepackage{amsmath}
\usepackage{cite}
\usepackage{amsfonts}
\usepackage{amssymb}
\usepackage{graphicx}
\usepackage{subfigure}
\usepackage{algorithm}
\usepackage{algpseudocode}
\usepackage{mathtools}

\DeclarePairedDelimiter\floor{\lfloor}{\rfloor}
\usepackage{textcomp}
\usepackage[export]{adjustbox}
\usepackage{float}
\usepackage{booktabs}
\usepackage{multicol}
\usepackage{xparse}
\usepackage{color,soul}
\usepackage[bottom]{footmisc}
\usepackage[binary-units=true]{siunitx}
\DeclareSIUnit{\belmilliwatt}{Bm}
\DeclareSIUnit{\dBm}{\deci\belmilliwatt}
\DeclareSIUnit[per-mode=symbol,per-symbol=p]{\Bps}{\byte\per\second}
\sethlcolor{yellow}
\usepackage{algpseudocode}
\makeatletter
\def\BState{\State\hskip-\ALG@thistlm}
\makeatother

\begin{document}

\title{Congestion-Aware Routing in Dynamic IoT Networks: A Reinforcement Learning Approach
}

\author{Hossam Farag and \v{C}edomir Stefanovi\'{c}\\
Department of Electronic Systems, Aalborg University, Denmark\\
	\{hmf,cs\}@es.aau.dk
}

	\maketitle
	\begin{abstract}
The innovative services empowered by the Internet of Things (IoT) require a seamless and reliable wireless infrastructure that enables communications within heterogeneous and dynamic low-power and lossy networks (LLNs). The Routing Protocol for LLNs (RPL) was designed to meet the communication requirements of a wide range of IoT application domains. However, a load balancing problem exists in RPL under heavy traffic-load scenarios, degrading the network performance in terms of delay and packet delivery. In this paper, we tackle the problem of load-balancing in RPL networks using a reinforcement-learning framework. The proposed method adopts Q-learning at each node to learn an optimal parent selection policy based on the dynamic network conditions. Each node maintains the routing information of its neighbours as Q-values that represent a composite routing cost as a function of the congestion level, the link-quality and the hop-distance. The Q-values are updated continuously exploiting the existing RPL signalling mechanism. The performance of the proposed approach is evaluated through extensive simulations and compared with the existing work to demonstrate its effectiveness. The results show that the proposed method substantially improves network performance in terms of packet delivery and average delay with a marginal increase in the signalling frequency. 
	\end{abstract}
	
\section{Introduction}

Internet of Things (IoT) represents a pervasive ecosystem of innovative services within different application areas~\cite{iot1}.
Low-power and Lossy Networks (LLNs) constitute the wireless infrastructure for a variety of IoT applications. The LLN is a set of interconnected resource-constrained devices, featuring limitations on power, processing and storage capabilities~\cite{LLN}.

The IPv6 Routing Protocol for LLNs (RPL)~\cite{RPL} was standardized as the de-facto routing protocol to meet the requirements of LLN applications. RPL organizes an IoT network as a Destination-Oriented Acyclic Graph (DODAG) rooted at the sink node, namely the DODAG root. The DODAG root represents the final destination for the traffic within the network level, bridging the topology with other IPv6 domains such as the Internet. Uplink routes are constructed and maintained using the periodic DODAG Information Object (DIO) messages~\cite{RPL}. RPL is a single path routing protocol where each node transmits its traffic directly to a preferred parent, which is exclusively selected based on the adopted Objective Function (OF)~\cite{suv}. Although RPL is originally designed for low traffic scenarios, performance issues arise under heavy traffic loads due to congestion, such as load imbalance, high packet loss and fast energy depletion of bottleneck nodes.  

Most of the proposed approaches to improve RPL against congestion are based on fixed strategies that incur expensive overhead~\cite{T-RPL}; as a result, the network does not work properly for dynamic IoT environments. In this respect, using machine learning frameworks could enable self-organizing operation of LLN devices in dynamic IoT environments \cite{ML1}. Reinforcement learning (RL) is a  machine learning technique capable of training an agent (IoT device) to interact intelligently and autonomously with an environment~\cite{RL}. Ultimately, adopting learning-based protocols to improve network performance can provide support for advanced IoT applications.
 
In this paper, we introduce an RL-based routing strategy to tackle the congestion problem in RPL networks. The proposed method is based on Q-learning where each node maintains a Q-table that represents the routing information of all its neighbours. The Q-values at each node are updated according to the received feedback function which is formulated as a composite routing metric that reflects the congestion level and link quality of each neighbour. Based on the received feedback, the node selects the neighbour with the minimum Q-value as its preferred parent. In order to comply with the standard RPL and keep the overhead to a minimum, the congestion information is embedded in the periodic DIO control messages. The period of DIO message exchanges is governed by the RPL Trickle timer~\cite{trickle}, whose reset strategy is modified to achieve a balance between overhead and fast dissemination of the feedback function. To the best of our knowledge, this is the first work that adopts machine learning to tackle the congestion problem in RPL-based networks. Performance evaluations and comparative analysis with respect to solutions available in the literature are carried out, showing that the proposed method achieves load balancing with improved performance in terms of packet delivery and average delay with a marginal increase in the signalling overhead.

The remainder of the text is organized as follows. Section~II presents the background and related work. Section~III introduces the load-balancing problem in RPL and the proposed Q-learning method. The performance evaluation is given in Section~IV, followed by the conclusions in Section~V. 

\section{Background and Related Work}
	\begin{figure}[t!]
		\centering
		\includegraphics[width=0.75 \linewidth]{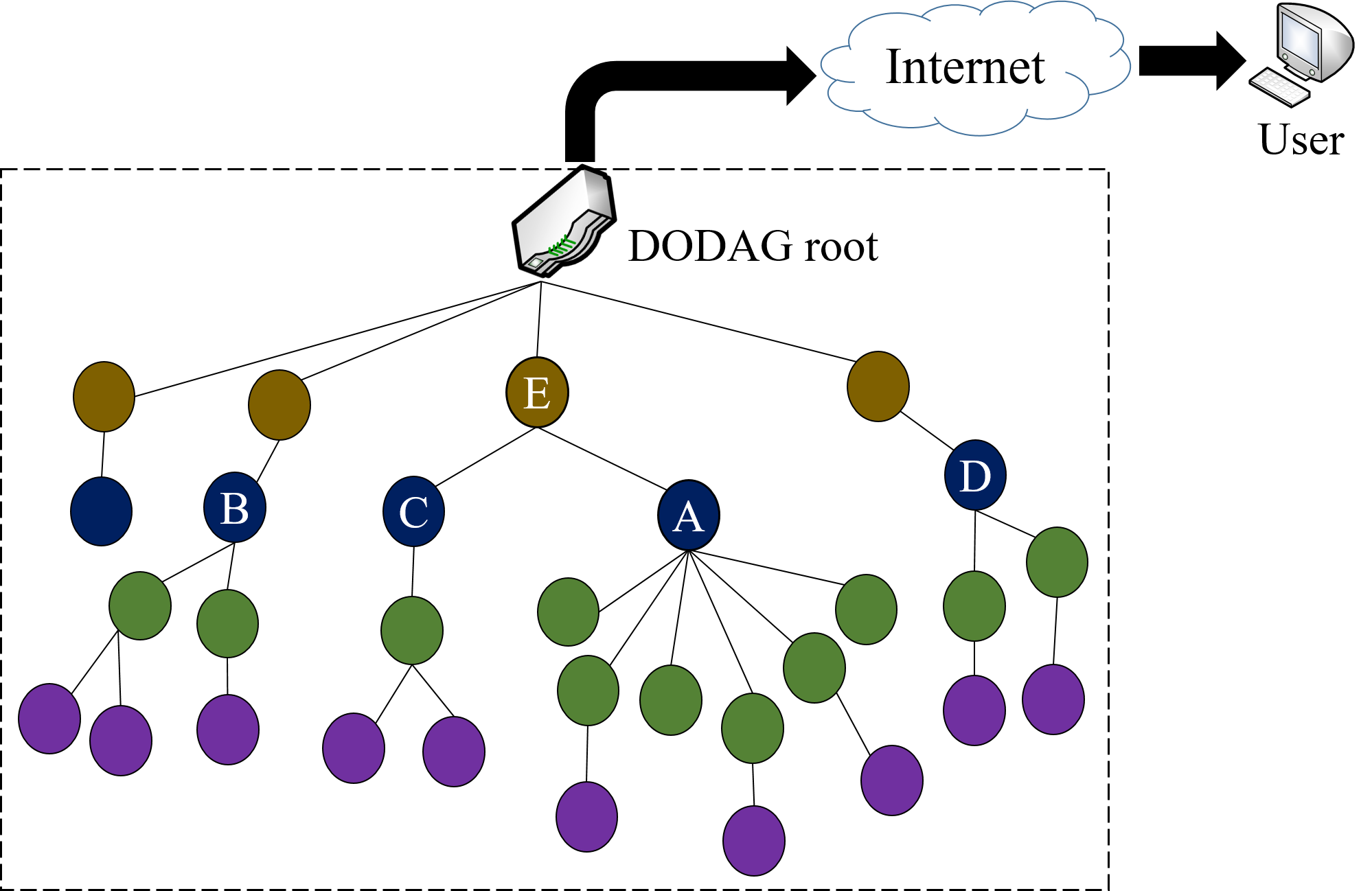}
		\caption{RPL network model. \label{f1}}
	\end{figure}
In RPL standard, the OF is used to describe the set of rules and policies that governs the process of route selection and optimization in a way that meets the different requirements of various applications. RPL decouples the route selection and optimization mechanisms from the core protocol specifications to enable its users to define routing strategies and local policies to fulfill the conflicting requirements of different LLN applications~\cite{RPL}. Currently, two OFs have been standardized for RPL, namely, the Objective Function Zero (OF0)~\cite{OF0} and the Minimum Rank with Hysteresis Objective Function (MRHOF)~\cite{MRHOF}. The two OFs rely solely on a single metric as the routing decision metric, hop-count in OF0 and Expected Transmission Count (ETX) in MRHOF. In heavy-traffic scenarios, LLN nodes are triggered to forward high amount of traffic which may cause congestion at particular parent nodes when the offered load exceeds the available queuing capacity~\cite{congs}. Consequently, the network suffers from consecutive packet delivery failures due to output queue overflows, also denoted as node congestion (henceforth, also simply referred to as congestion). Node congestion occurs even in the case of low-rate traffic because nodes near to the DODAG root have to relay high amount of traffic. Therefore, node congestion and imbalanced routing tree topology have a profound impact on the network performance in terms of packet loss, delay, and energy consumption. In RPL specifications~\cite{RPL}, there is no explicit mechanism to detect and react to node congestion. Instead, all traffic will be forwarded through the selected preferred parent as long as it is reachable, without any attempt to perform load balancing, which ultimately leads to poor quality of service (QoS). Therefore, incorporating load-balancing mechanism in RPL routing is essential to maintain efficient network performance in terms of delay and reliability.

Several rule-based approaches have been proposed in the literature to tackle the congestion problem in RPL networks. The authors in~\cite{C0} developed an OF using fuzzy logic to improve packet delivery and energy consumption. The method utilizes the hop-count, ETX and residual energy to select the best parent, however, it does not consider the congestion level in the selection process. In~\cite{C1}, the authors proposed a non-cooperative game theory-based solution to mitigate node congestion in RPL, which attempts to find the optimal sending rate of each node based on the information of buffer loss and channel loss. Besides the increased overhead, adjusting the sending rate may violate the potential requirement of IoT applications where a fixed refresh rate is required. A context-aware OF was proposed in~\cite{C2} to mitigate node congestion and improve network lifetime under heavy traffic. The method considers the residual energy and queue utilization when selecting the preferred parent, however, it suffers from increased overhead when distributing the routing information among neighbour nodes. A load-balancing parent selection was proposed in~\cite{C3}, where all feasible parents compete for packet forwarding when a node transmits a packet. The method incurs a significant energy waste when all the feasible parents wake up their radio, and also does not guarantee that the winner is the one with best link quality.

A number of machine learning-based protocols was introduced to improve routing in next-generation networks~\cite{ML2, ML3, ML4}, however, these works are not compliant with the RPL specification. The authors in \cite{ML5} introduced iCPLA, a cross-layer optimization method using reinforcement learning to improve RPL performance under heterogeneous traffic. The proposed method utilizes the collision probability to select the optimal parent. However, the method considers only the channel congestion due to collisions and neglects node congestion which has a crucial impact on the network performance, as shown in the next section.    
	\begin{figure}[t!]
		\centering
		\includegraphics[width=0.7 \linewidth]{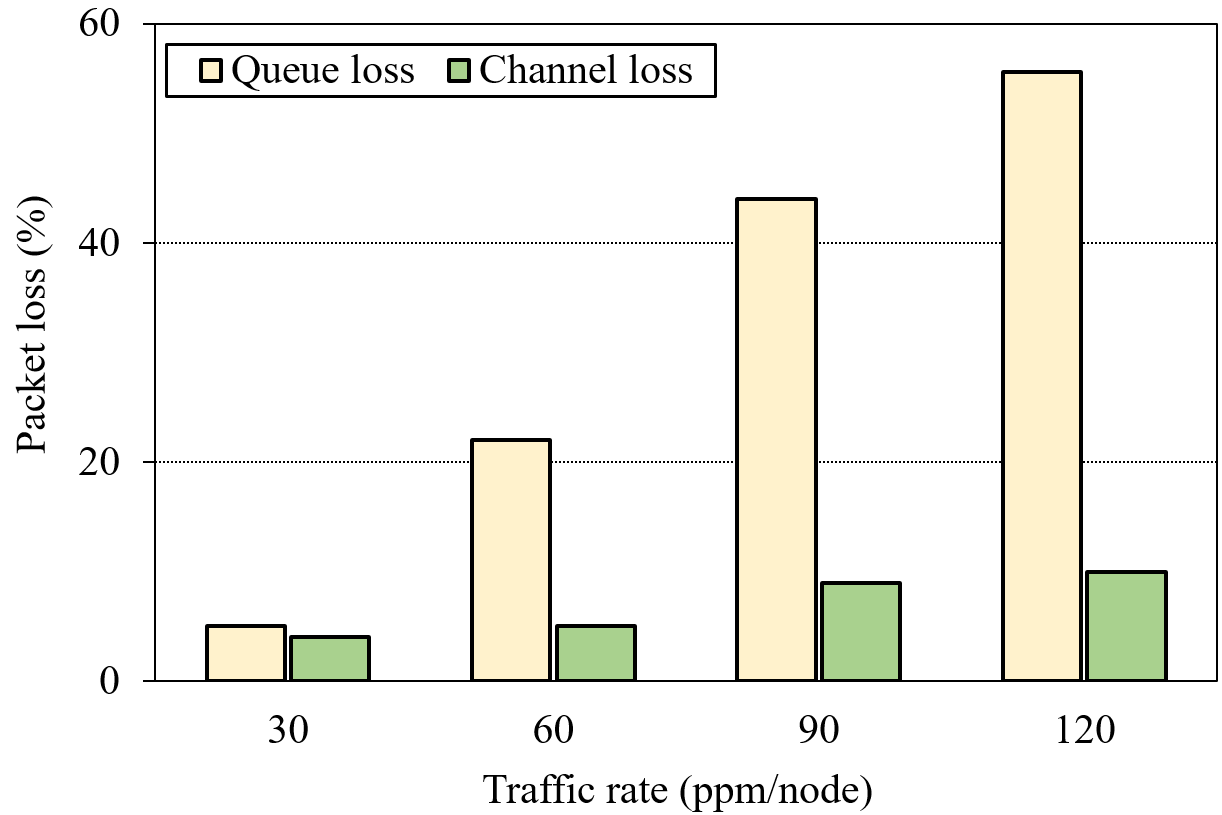}
		\caption{Packet delivery performance under varying traffic load. \label{PDR1}}
	\end{figure}
\section{The Proposed  Method}

First, we emphasize the load balancing problem in the standard RPL. Then, we introduce the proposed Q-learning method to mitigate node congestion and achieve load balancing.

\subsection{Load-Balancing Problem in RPL}
\label{sec:LB}

We consider a typical LLN model in an IoT application, e.g., industrial automation as depicted in Fig.~\ref{f1}. In this model, a set of sensor nodes are distributed to form the LLN that is rooted at the DODAG root (border router). The DODAG root connects the LLN to the public Internet or a private IP-based network. The nodes utilize IEEE~802.15.4 links to communicate with each other, and use RPL to construct routes towards the DODAG root. We first investigate the packet delivery performance towards the DODAG root of an LLN consisting of 30~nodes via MATLAB simulations. The LLN structure shown in Fig.~\ref{f1} depicts a characteristic snapshot of the RPL topology using MRHOF at the end of a simulation. Fig.~\ref{PDR1} shows the percentage of packets lost at the link layer and also at the output queue under different traffic rates with each node having a buffer size of 10~packets; note that the assumed traffic-load range covers values that can be expected in practice. The results shown in Fig.~\ref{PDR1} represents the average over all nodes in the network. The figure demonstrates that, as the traffic rate increases, the losses due to the node congestion tend to increasingly dominate over the channel losses, quickly becoming the main reason of packet delivery degradation.
	\begin{figure}[t!]
		\centering
		\includegraphics[width=0.7 \linewidth]{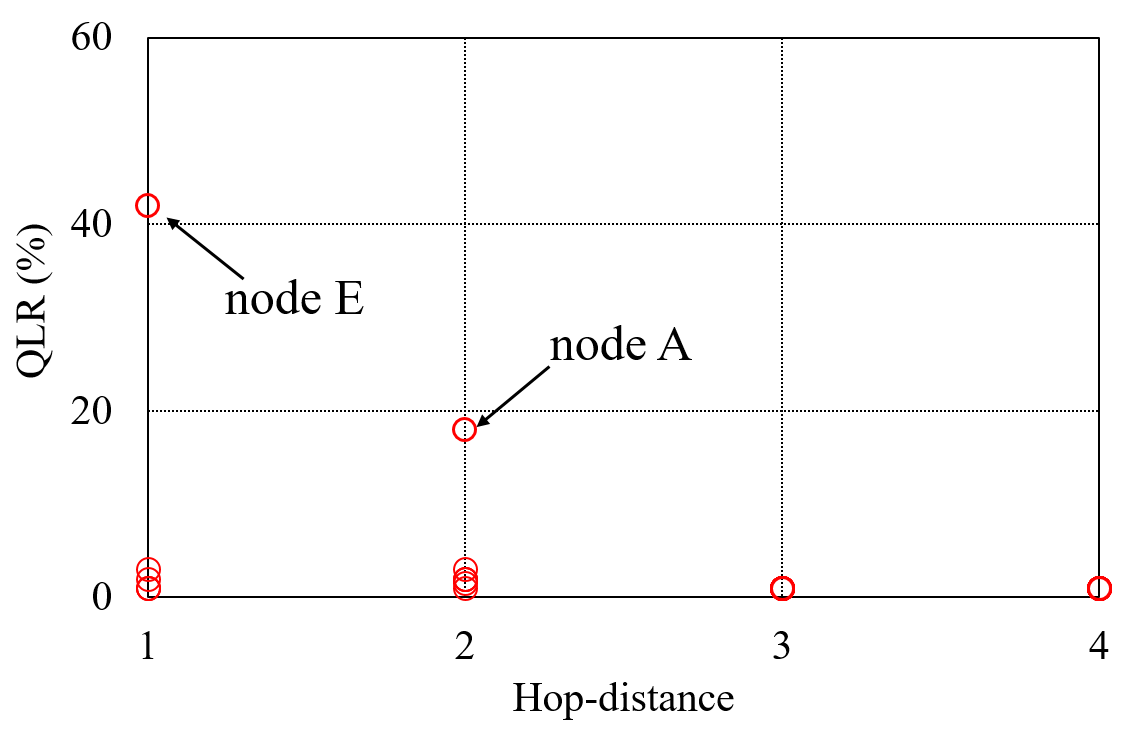}
		\caption{Queue losses at each node at 90 ppm/node rate. \label{queue1}}
	\end{figure}

Fig.~\ref{queue1} shows the Queue Loss Ratio (QLR) of each node according to its hop-distance from the DODAG root. The QLR is calculated as the ratio between the packets dropped due to buffer overflow and the total transmitted packets. Initially, it might be intuitive that nodes that are one-hop distance from the DODAG root experience the highest queue loss levels. However, the results presented in Fig.~\ref{queue1} (corresponding to the RPL topology in Fig.~\ref{f1}) reveal that, under heavy load, there is an imbalance among queue losses among nodes at the same hop-distance from the root. For instance, node~$A$ has the highest QLR among five nodes that are 2 hops from the root. Similarly, node $E$ has the highest QLR among four nodes that are a single hop from the root. This is due to the inefficient parent selection mechanism in RPL, which leads to imbalanced sub-tree sizes among nodes at the same hop-distance. Since LLN nodes have small queue sizes, typically 4 to 10 packets, their queues start to overflow before the congestion level is heavy enough to be detected through the ETX parameter. Therefore, children nodes are not aware of the backlog status of their parents, continuing to transmit packets even if the parents suffer from consecutive queue losses. In summary, there is a need for an efficient load-balancing strategy in RPL.      

\subsection{The Proposed Q-learning for Load-balancing in RPL}

Q-learning is a model-free RL technique, where the agent learns to select the best action according to the maintained Q-table~\cite{Q-learning}. In our approach, the Q-table corresponds to the routing table at each node (agent) and the action represents the selection of the preferred parent. The learning goal is to find the optimal parent selection policy according to the network dynamics, e.g congestion level and link quality. 

Each node $x$ maintains a candidate neighbour set $N(x)$ which includes all nodes that are within the communication range of $x$.
Denote by $ Q_x(y)$ a Q-value that is the preference of node $x$ to select node $y$ as its parent, where $y\in N(x)$.
The Q-values are periodically updated in the following way
\begin{equation} \label{upd}
    Q_x^{\text{new}}(y)=Q_x^{\text{old}}(y)+\alpha\left[R(y)-Q_x^{\text{old}}(y)\right]
\end{equation}
where $Q_x^{\text{new}}$ and $ Q_x^\text{old}(y)$ are the Q-values for the current and the previous intervals, respectively; the interval duration is determined by the Trickle timer, as will be elaborated later. $\alpha$ is the learning rate, and $R(y)$ is the estimate of the feedback function, i.e., the reinforcement signal. Note that the Q-values are initialized to zeros for all nodes upon network deployment. The value $ Q_x(y)$ represents the routing metric of the path between $x$ and $y$. The main idea of the proposed method is to enable the nodes to learn the congestion levels at their neighbours through the feedback $R(y)$, and utilize this information to select the best parent in order to achieve load-balancing. We represent the congestion level at node $y$ through the Backlog Factor $\mathrm{BF}(y)$, which denotes the ratio between the current queue length and the total queue size. An exponentially weighted moving average filter is applied for calculation
of $\mathrm{BF}(y)$. The congestion status $\mathrm{BF}(y)$ is distributed to all nodes in the set $N(y)$, as illustrated later. To obtain the optimal network performance, in heavy traffic mode, the node learns to select the parent that is less congested, while in light traffic mode, i.e., no congestion, the node learns to select the parent with the best link quality and the shortest hop-distance.  Thus, we formulate $R(y)$ as a composite routing metric as follows
\begin{equation} \label{rw}
    R(y)=\lambda(y)\mathrm{BF}(y)+\mathrm{ETX}(x,y)+H(y)
\end{equation}
where $H(y)$ denotes the hop-count of node $y$ towards the DODAG root, and $\mathrm{ETX}(x,y)$ is the ETX measurement between $x$ and $y$.\footnote{More complex formulations for $R(y)$ can be devised. However, the simple metric in \eqref{rw} fosters an excellent performance, as shown in Section~\ref{sec:evaluation}.}
In RPL standard, $\mathrm{ETX}(x,y)$ is exchanged periodically, and is calculated as~\cite{etx} 
\begin{equation}
\mathrm{ETX}(x,y)=\frac{\#\,\textrm{of total transmissions from\,}x\,\textrm{to}\, \,y}{\# \,\textrm{of successful transmissions from}\,x\,\textrm{to}\,\,y}.
\end{equation}
The parameter $\lambda(y)$ is used to control the weight of $\mathrm{BF}(y)$ in \eqref{rw}, such that it reflects congestion level of the node $y$. 
Since it holds that $0\leq \mathrm{BF}(y)\leq1$, we define $\lambda(y)$ as
\begin{equation} \label{th}
\lambda(y) = \max\left(\frac{\mathrm{BF}(y)}{\mathrm{BF}_{\text{th}}}, 1-\frac{\mathrm{BF}(y)}{\mathrm{BF}_\text{th}}\right)
\end{equation}
where $\mathrm{BF}_\text{th}$ is a design parameter that represents the threshold after which the parent node is assumed to be congested. That way, in congestion situations, i.e., when $\mathrm{BF}(y) > \mathrm{BF}_\text{th}$, $\mathrm{BF}(y)$ will have a noticeable effect in \eqref{rw} compared to $\mathrm{ETX}(x,y)$ and $H(y)$. Conversely, in light traffic scenarios, when $\mathrm{BF}(y) < \mathrm{BF}_\text{th}$, the feedback function is influenced more by $\mathrm{ETX}(x,y)$ and $H(y)$ compared to $\mathrm{BF}(y)$. Based on the received feedback $R(y)$, the node updates the value of $Q_x(y)$ according to \eqref{upd}. 

	\begin{figure}[t!]
		\centering
		\includegraphics[width=0.5 \linewidth]{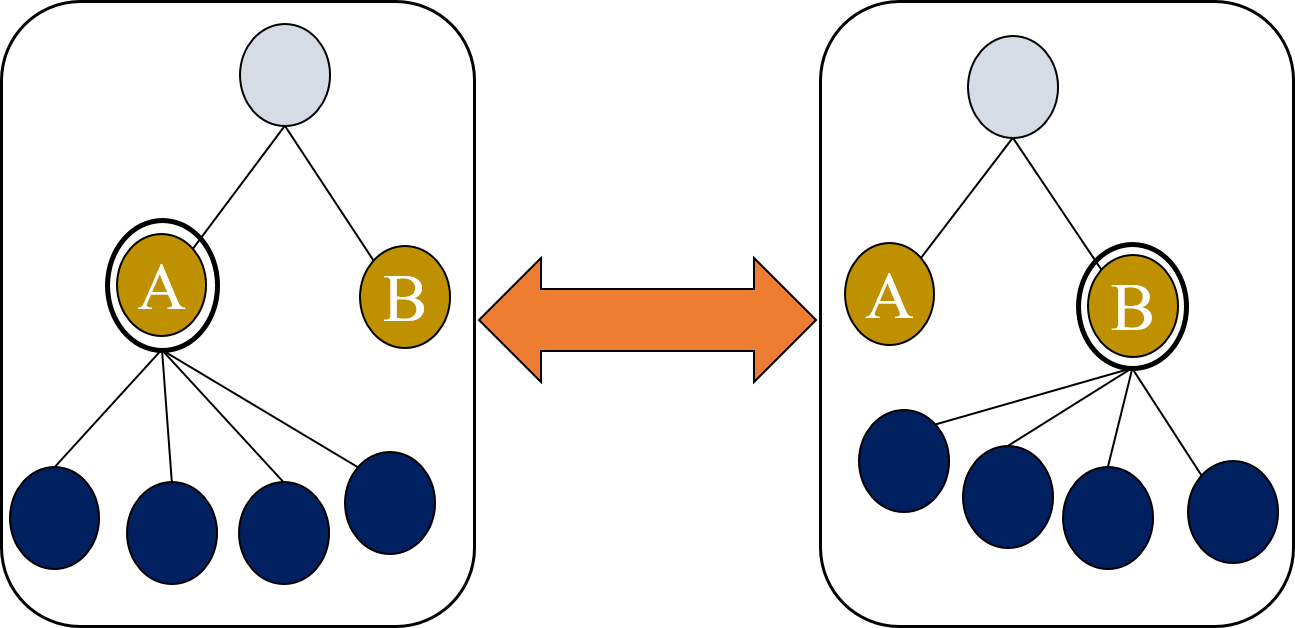}
		\caption{Illustration of the thundering herd problem. \label{thund}}
	\end{figure}

A straightforward approach would be to use the greedy policy and select the neighboring node with the minimum Q-value as its preferred parent.
However, the approach in which all nodes utilize only the exploitation phase that follows the greedy policy may lead to the thundering herd problem~\cite{thunder}, illustrated in Fig.~\ref{thund}. The four leaf nodes in Fig.~\ref{thund} may select node $A$ as their preferred parent as it has the minimum Q-value, which may incur congestion. Upon detecting congestion and selecting a new preferred parent, the four nodes may change simultaneously to the same parent node $B$ causing congestion again; such cycle may be repeated indefinitely without achieving load balancing. To overcome this problem, the nodes have to explore other alternatives using a certain probability. Specifically, we refer to $P_x(y)$ as the probability of node $x$ selecting node $y$ as the preferred parent, given by
\begin{equation}
P_x ( y )= 1-\frac{e^{Q_x(y)/\theta}}{\sum_{k\in N(y)} e^{Q_x(k)/\theta}}
\label{prob}
\end{equation}
where $\theta$ is a parameter that determines the amount of exploration. The probabilistic parent selection in \eqref{prob} represents a modified version of the softmax action selection strategy. In other words, it implies that a neighbour with low Q-value will be most likely to be selected as a preferred parent, while other neighbours with higher Q-values will be selected with smaller probabilities. Accordingly, the nodes will obtain an efficient experience and avoid thundering herd problem as well.

The congestion level $\mathrm{BF}(y)$ needs to be distributed to the set $N(y)$ in order to make the right action according to the network status.
In RPL standard, DIO messages are periodically exchanged between neighbour nodes and carry the RANK of the transmitted node, i.e, $H(y)$.
In our proposed approach, we implicitly embed $\mathrm{BF}(y)$ into the RANK value in the transmitted DIO message as follows
\begin{equation}
\mathrm{RANK}^\text{new}(y)= \eta \left(H(y)+1\right)+(\eta-1) \mathrm{BF}(y), 
\label{rank}
\end{equation}
where $\eta$ is a positive integer that enables decoding two values ($\mathrm{BF}(y)$ and $H(y)$) from a single numeric field ($\mathrm{RANK}^\text{new}$). The value of $\eta$ should be selected in a way to keep $\mathrm{RANK}^\text{new}(y)$ within its 16~bit boundary~\cite{RPL}. Specifically, a neighbor node that receives the DIO message from $y$ extracts the two values $\mathrm{BF}(y)$ and $H(y)$ from $\mathrm{RANK}^\text{new}(y)$ as
\begin{equation}
\begin{split}
 \mathrm{BF}(y)&=\frac{\text{mod}\left(\mathrm{RANK}^\text{new}(y),\eta\right)}{\eta-1}\\
 H(y)&=\floor*{\frac{\mathrm{RANK}^\text{new}(y)}{\eta}}-1
\end{split}
\label{decode}
\end{equation}
where $\text{mod}()$ is the modulo operation. Therefore, in the proposed method, the congestion information are distributed among neighbour nodes without changing the DIO message format nor adding any additional control message overhead.  
	\begin{figure}[t!]
		\centering
		\includegraphics[width=0.8 \linewidth]{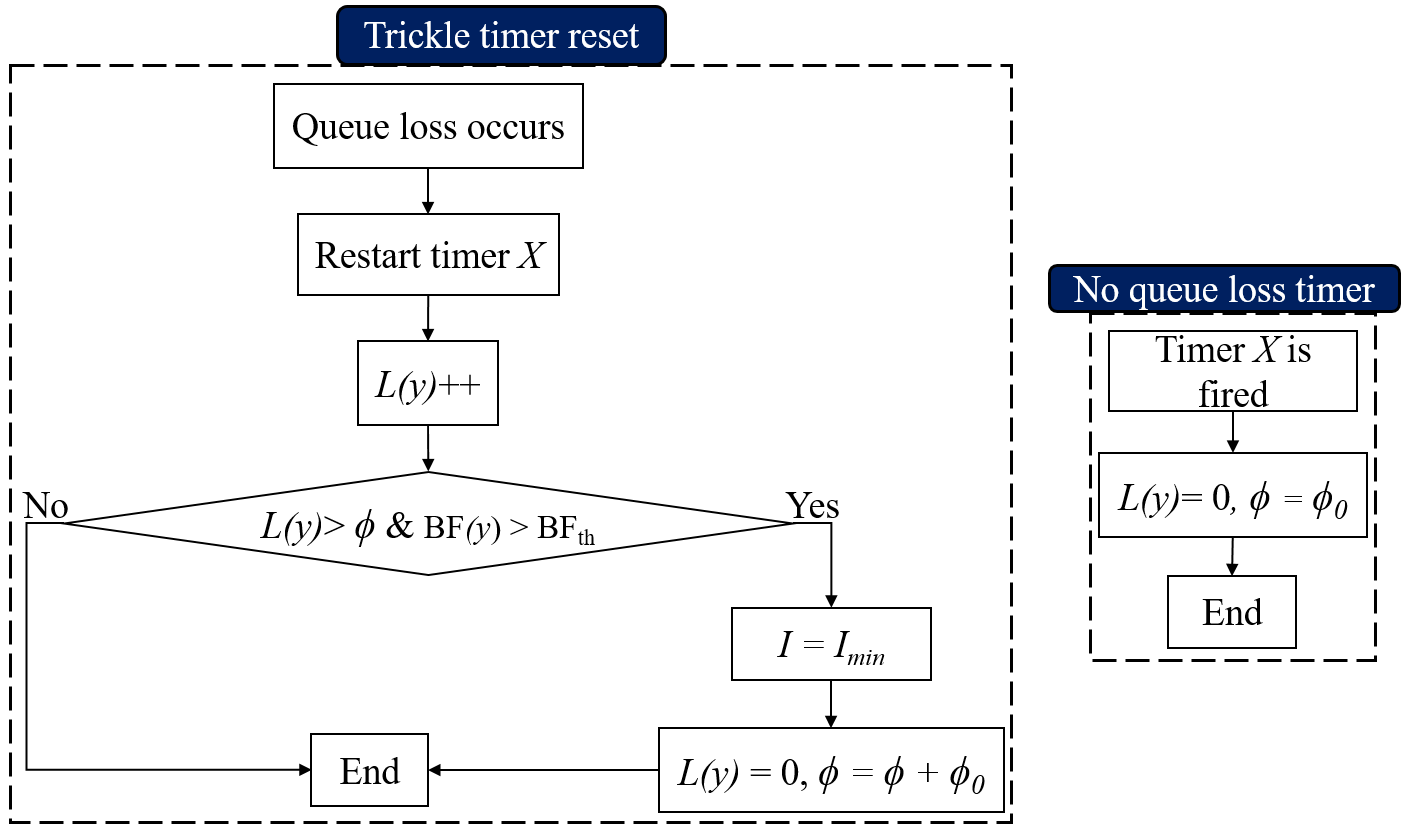}
		\caption{The proposed Trickle timer reset strategy. \label{trickle}}
	\end{figure}

The transmission interval of the DIO messages is controlled by the Trickle timer algorithm~\cite{trickle}. According to the standard, the Trickle timer is reset to its minimum interval $I_{\min}$ only when changes occur in the network topology. As long as the network is consistent, the DIO message interval is doubled up to a certain maximum value. With such strategy, the nodes may have inaccurate and outdated congestion information from neighbor nodes. On the other hand, frequent resetting of the Trickle timer may increase the routing overhead. To tackle this conflict, we introduce a modified reset strategy to the Trickle timer. The proposed strategy is shown in Fig.~\ref{trickle} and is described as follows. A node resets its Trickle timer to $I_{\min}$ when it experiences a certain number $\phi$ of consecutive queue losses. Specifically, the small queues of LLN nodes my fill up temporarily even when there is no congestion, that often results in false positive for congestion if it is declared too early, which in turn incurs unnecessary DIO overhead. For this reason, we use consecutive queue losses to reset the Trickle timer. After resetting the timer to $I_{\min}$, the value of $\phi$ is increased by a fixed value $\phi_0$  to limit the reset frequency and the DIO overhead. If no queue losses are detected within a timer $X$, the algorithm values are reinitialized as shown in Fig.~\ref{trickle}.  
\section{Evaluation}
\label{sec:evaluation}

\begin{table}[t!]
		\centering
		\caption{Evaluation parameters.}
		\label{t1}
		\begin{tabular}{ll}
			\toprule
			Parameter & Value \\
				\midrule
			Network size & \num{30} nodes\\
			Packet length & \SI{100}{\byte} \\
			Traffic load per node & \num{30}, \num{60}, \num{90}, \num{120} ppm \\
			Propagation model & Shadowing (Log-normal)\\
			Standard deviation & \SI{14}{\dB} \\
			PHY and MAC protocol & IEEE 802.15.4 with CSMA/CA\\
			Slotframe length & \num{500} slots\\
			Time slot duration &  \SI{10}{\milli\second} \\
			No. of retransmissions & \num{3}\\
			$\alpha$ & \num{0.3}\\
			$\mathrm{BF}_\text{th}$ & \num{0.5}\\
			$\eta$ & \num{100}\\
			$\phi_0$ & \num{2}\\
			Timer $X$ & \SI{100}{\milli\second}\\
			$I_{min}$ &  \SI{3}{\second}\\
			\bottomrule
		\end{tabular}	
	\end{table}

 We consider a typical IoT network in a monitoring scenario, where a set of 30 randomly placed nodes are reporting to a single DODAG~root.
The nodes communicate using IEEE~802.15.4 PHY layer and CSMA/CA for channel access. Each node generates a 100-byte packet following a Poisson arrival model and forwards it to its preferred parent; the traffic arrival intensity per node (the traffic load) is a parameter whose values are given in Table~\ref{t1}. We consider a log-normal shadowing propagation model with the standard deviation specified in Table~\ref{t1}~\cite{std}. 
The table also lists the values of other relevant parameters.\footnote{The values of the parameters $\mathrm{BF_{th}}$, $\phi_0$ and $X$ were found via optimization and are independent of the traffic load. Our investigations also showed that these values are robust to the change in the number of network nodes.} To verify the effectiveness of the proposed method, we compare its performance with iCPLA \cite{ML5} and RPL using MRHOF (RPL-MRHOF)~\cite{MRHOF}. 
The evaluation was performed using Monte Carlo simulations in MATLAB; the presented results are the averages of 10~simulation runs for each value of the traffic load, where each run lasted for a duration of 1000~consecutive slotframes.

\begin{figure}[t!]
		\centering
		\includegraphics[width=1 \columnwidth]{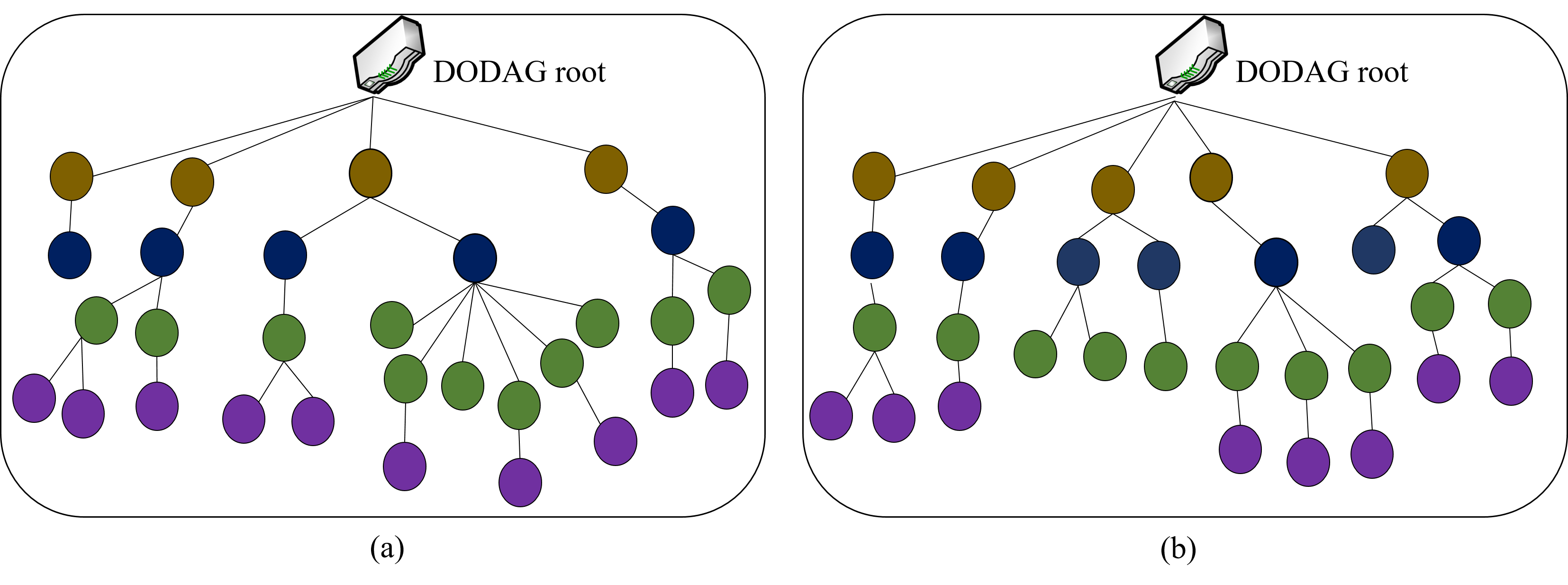}
		\caption{Routing topology of: (a) RPL-MRHOF; (b) Proposed method. \label{topp}}
	\end{figure}

Fig.~\ref{topp} shows a sample of the routing topology of both RPL-MRHOF and the proposed method. Obviously, the proposed method achieves load balancing between intermediate nodes, as the nodes learn to avoid congested nodes when selecting their parents. Quantitatively, the method reduces the standard deviation of the number of children per node from~1.3 to 0.75.

\begin{figure}[t!]
		\centering
		\includegraphics[width=0.8 \linewidth]{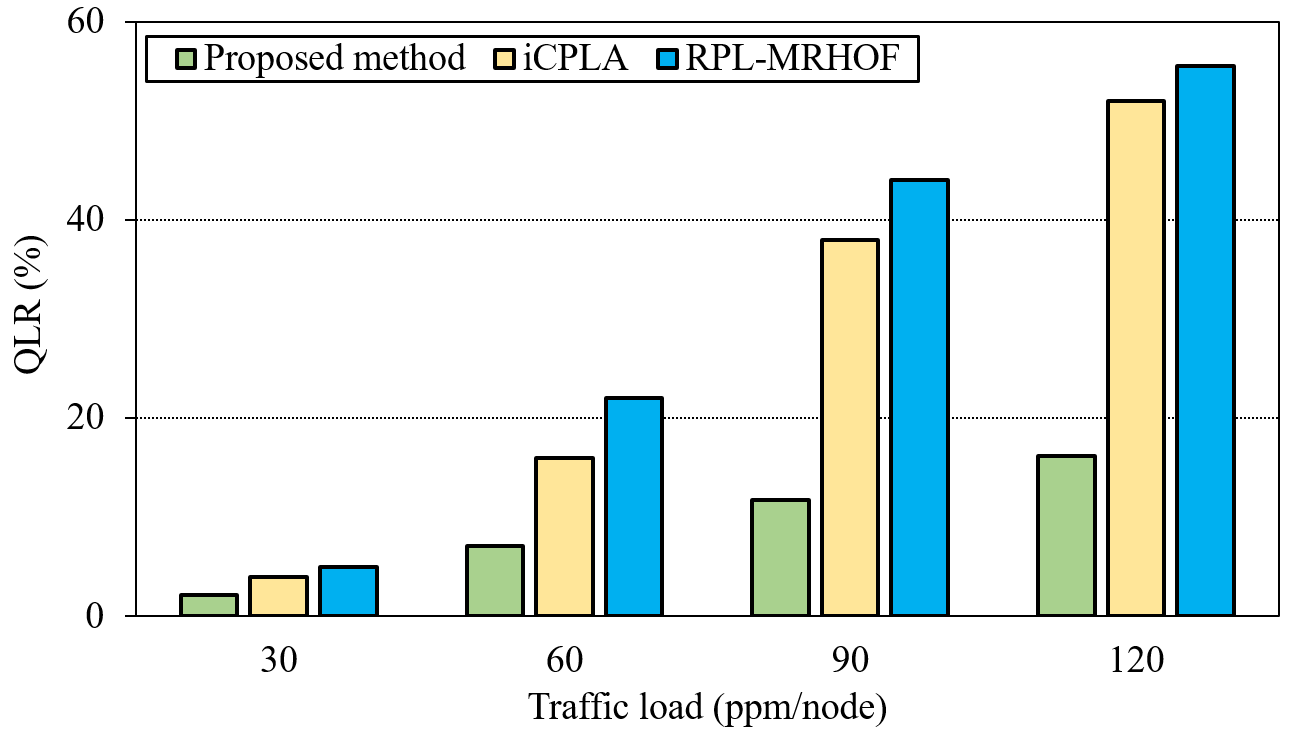}
		\caption{QLR comparison with varying traffic load. \label{QLR}}
	\end{figure}

Fig.~\ref{QLR} presents a comparison of the average QLR with varying traffic load: While iCPLA fares better than RPL-MRHOF, the proposed Q-learning-based method achieves the lowest QLR due to a more balanced load distribution 
For instance, the proposed method reduces the QLR of RPL-MRHOF by 71\% at 120 ppm/node, while iCPLA achieves only 5\% reduction at the same traffic load.
Specifically, the parent selection strategy in iCPLA considers only the link-level information, i.e, collision probability, which is insufficient to solve node congestion under heavy load. 

	\begin{figure}[t!]
		\centering
		\includegraphics[width=0.8 \linewidth]{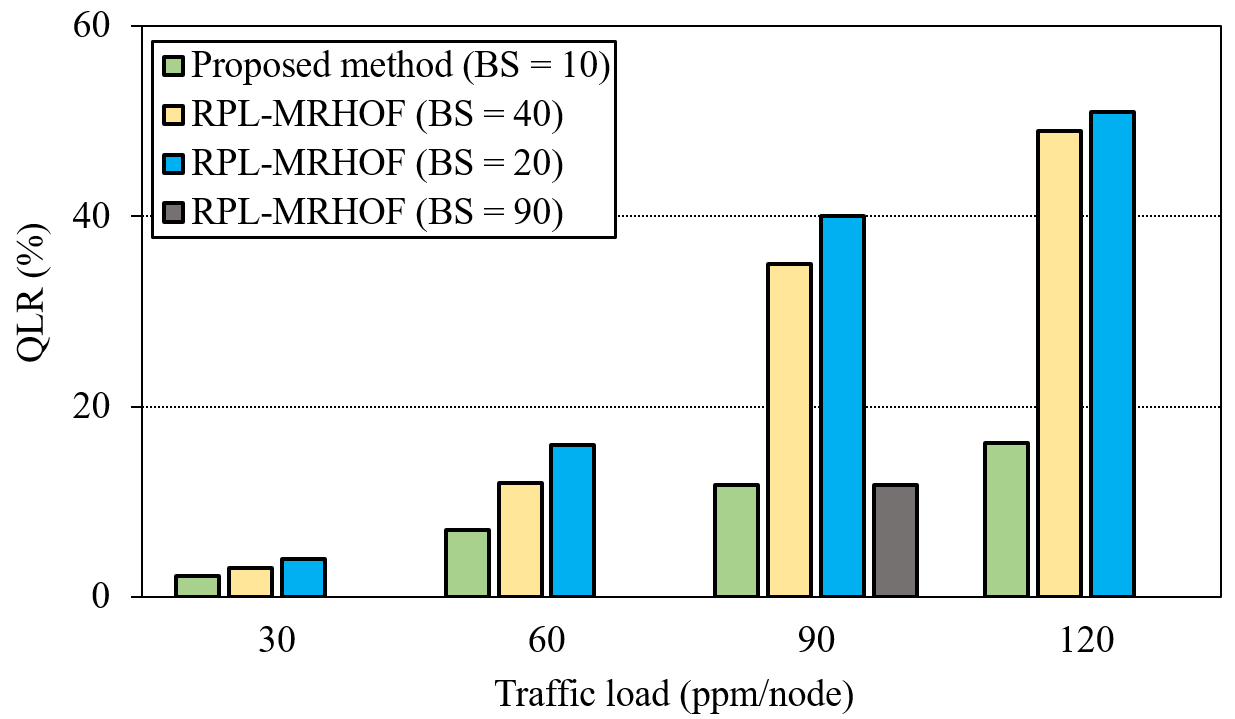}
		\caption{Effect of buffer size on QLR. \label{QLRB}}
	\end{figure}

As already mentioned, LLN devices have small queue sizes that start to overflow quickly as the traffic load increases.
In this respect, a solution could be to increase the Buffer Size (BS) to avoid node congestion.
However, as shown by Fig.~\ref{QLRB}, increasing the BS has a marginal effect to mitigate queue losses, especially under heavy load scenarios. Specifically, increasing the BS of RPL-MRHOF from 20 to 40~packets reduces the QLR by 12\% and 3\% at traffic loads of 90 ppm/node and 120 ppm/node, respectively.
The figure also shows that, in order for RPL-MRHOF to match QLR performance of the proposed method when the traffic load is 90 ppm/node, the BS should be 90~packets, which is infeasible in practice for the resource-constrained LLN nodes. In summary, achieving a balanced routing topology has a more significant impact on node congestion mitigation than increasing the BS.
	\begin{figure}[t!]
		\centering
		\includegraphics[width=0.8 \linewidth]{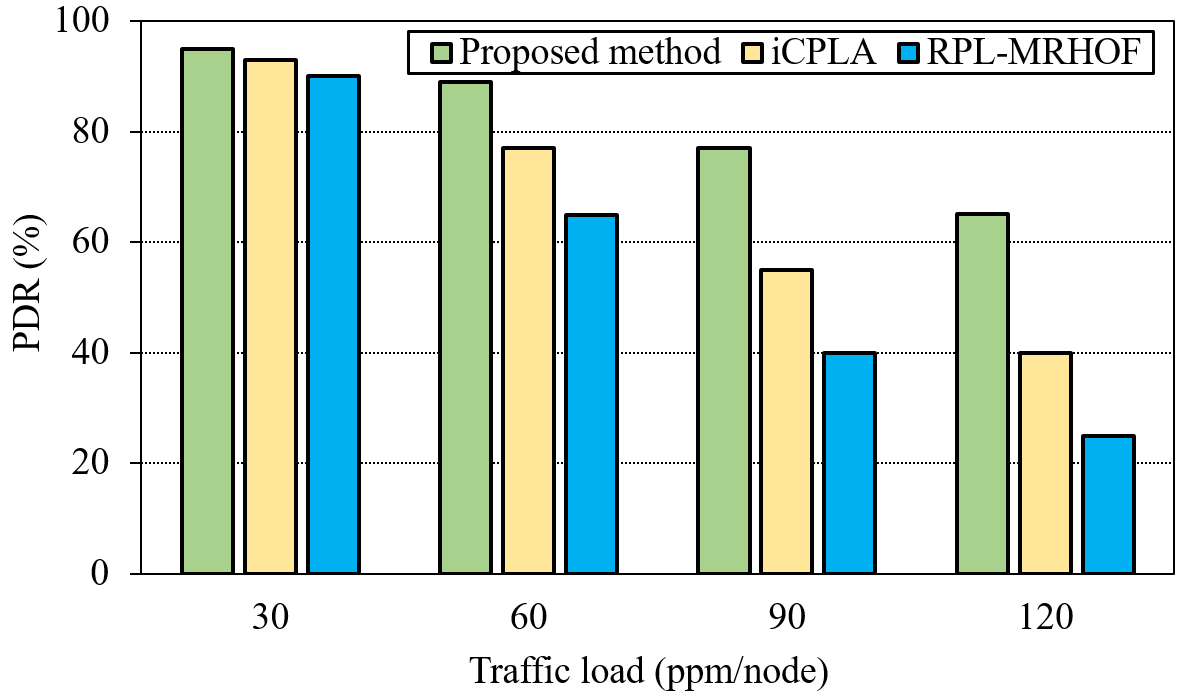}
		\caption{PDR comparison with varying traffic load. \label{PDR}}
	\end{figure}

As noted in Section~\ref{sec:LB}, queue losses are the main reason for the packet delivery degradation in the network. Therefore, the reduction in QLR is directly translated to improved packet delivery performance, as shown in Fig.~\ref{PDR}. The figure depicts the Packet Delivery Ratio (PDR), which is the number of packets successfully delivered to the DODAG root divided by the total generated packets. At 120 ppm/node, the proposed method enhances the PDR performance of RPL-MRHOF by 160\% compared to a 60\% improvement achieved by iCPLA. 
	\begin{figure}[t!]
		\centering
		\includegraphics[width=0.85 \linewidth]{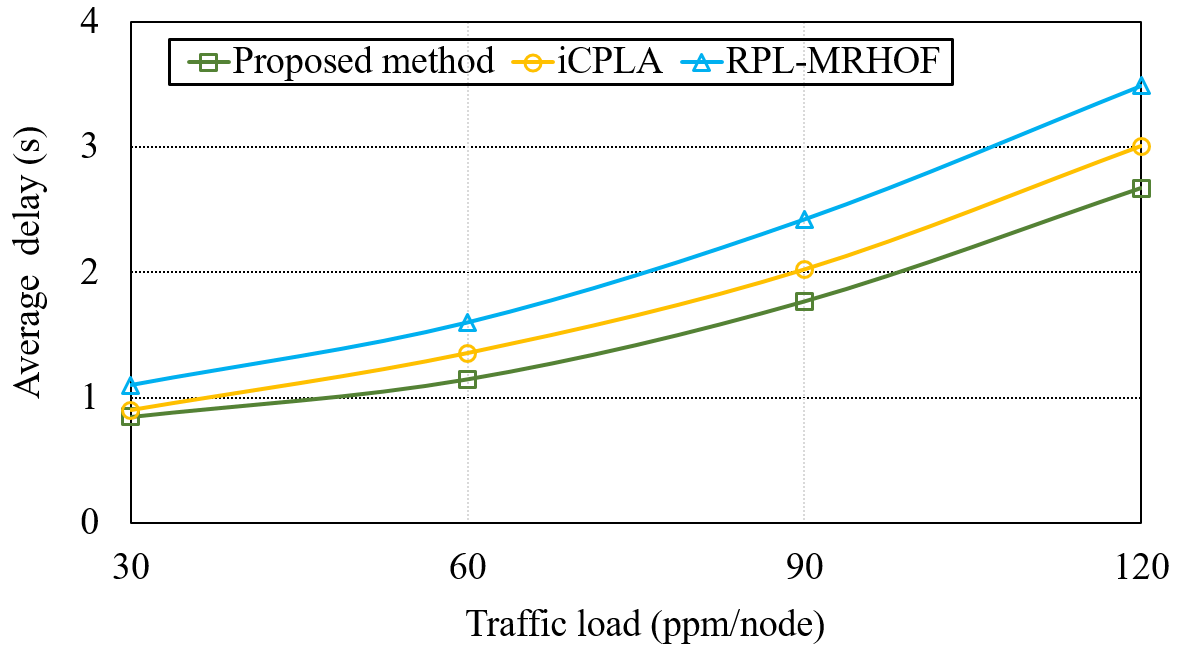}
		\caption{Average delay comparison with varying traffic load. \label{delay}}
	\end{figure}

Fig.~\ref{delay} compares the average delay of successfully received packets by the DODAG root under varying traffic load.
As the traffic load increases, the average delay in all three methods increases as more packets are backlogged and a packet has to spend more time in the output queue before transmission. However, the proposed method guides the nodes to a fair queue utilization strategy that helps to reduce the average backlogged packets per node. Accordingly, the queuing time is decreased which reduces the average delay in turn. For instance, at 90~ppm/node, the proposed method reduces the average delay of RPL-MRHOF and iCPLA by 27\% and 12\%, respectively. In other words, the proposed method outperforms the competitors with a significantly higher packet delivery ratio, and a lower delay of the delivered packets.

\begin{figure}[t!]
		\centering
		\includegraphics[width=0.8 \linewidth]{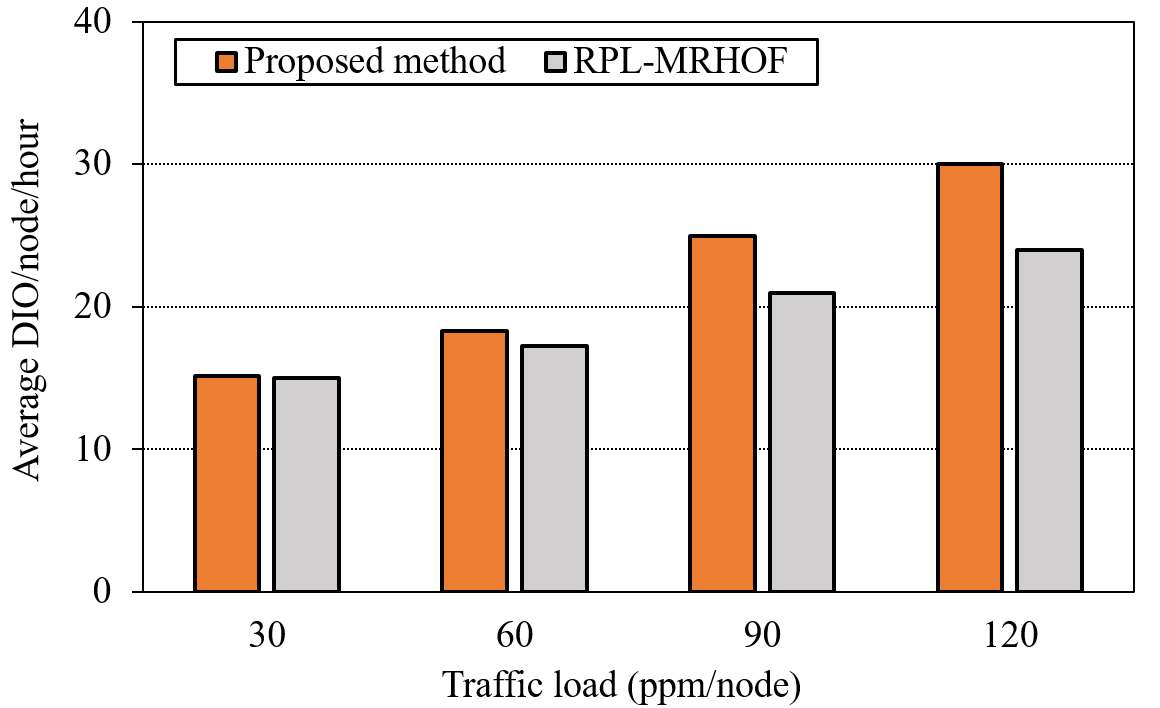}
		\caption{Average DIO overhead with varying traffic load. \label{dio}}
	\end{figure}

Fig.~\ref{dio} depicts the average number of transmitted DIO messages of each node under different traffic loads for the proposed method and RPL-MRHOF. As the figure demonstrates, the proposed method incurs higher DIO overhead compared to RPL-MRHOF, especially under heavy load. This is because our proposed method  resets the Trickle timer more frequently to fast distribute the feedback function when a node suffers from congestion. However, under heavy load scenarios, the increased overhead is still insignificant compared to the total traffic in the network. For instance, at 90 ppm/node, the DIO overhead represents less than 0.5\% of the total traffic. Moreover, the increased DIO overhead could be a reasonable cost to the obtained performance improvements in the PDR and the average delay.    

\section{Conclusion}
In this paper, we have proposed a RL-based routing approach to mitigate node congestion and improve routing performance of RPL networks. Each node adopts Q-learning to select it preferred parent based on a composite feedback function. The congestion level at each node is embedded in the RANK metric and distributed using a modified Trickle timer strategy. Performance evaluations have been carried out and proved the effectiveness of our proposed method to achieve load balancing and improve the network performance in terms of packet delivery and average delay.  


\begin{thebibliography}{00}

	\bibitem{iot1} H. Tran-Dang, N. Krommenacker, P. Charpentier and D. Ki, ``Toward the Internet of Things for Physical Internet: Perspectives and Challenges,"  \textit{IEEE Internet of Things J.}, vol. 7, no. 6, pp.~4711-4736, Jun. 2020.

	
	 \bibitem{LLN} B. Ghaleb, \textit{et al.}, ``A survey of limitations and enhancements of the ipv6 routing protocol for low-power and lossy networks: A focus on core operations," \textit{IEEE Commun. Surv. Tut.}, vol. 21, no. 2, pp. 1607-1635, Secondquarter 2019.
	 
	 \bibitem{RPL}  T. Winter \textit{et al.}, \textit{RPL: IPv6 routing protocol for low-power and lossy networks}, IETF RFC 6550, Mar. 2012.
	 
	 \bibitem{suv} B. Ghaleb, \textit{et al.}, ``A survey of limitations and enhancements of the ipv6 routing protocol for low-power and lossy networks: A focus on core operations," \textit{IEEE Commun. Surv. Tut.}, vol. 21, no. 2, pp. 1607-1635, Second Quarter 2019.
	 
	 \bibitem{T-RPL} J. V. V. Sobral, \textit{et al.}, ``Routing protocols for low power and lossy networks in internet of things applications," \textit{Sensors}, vol. 19, no. 9, pp. 1-40, May 2019.
	 
	 \bibitem{ML1} S. K. Sharma and X. Wang, ``Toward massive machine type communications in ultra-dense cellular iot networks: Current issues and machine learning-assisted solutions," \textit{IEEE Commun. Surv. Tut.}, vol. 22, no. 1, pp. 426-471, Firstquarter 2020.
	     
	  \bibitem{RL} D. Praveen Kumar, T. Amgoth and C. S. R. Annavarapu, ``Machine learning algorithms for wireless sensor networks: A survey," \textit{Information Fusion}, vol. 49, pp. 1-25, Sept. 2019.
	  
	  \bibitem{trickle} P. Levis \textit{et al.}, \textit{The Trickle Algorithm}, [Online]. Available: https://tools.ietf.org/html/rfc6206.
	 	
	  \bibitem{OF0} P. Thubert, \textit{Objective function zero for the routing protocol for low-power and lossy networks (rpl)}, IETF RFC 6552, Mar. 2012.
	 
	  \bibitem{MRHOF} O. Gnawali and P. Levis, \textit{The minimum rank with hysteresis objective function}, IETF RFC 6719, Sept. 2012.
	   
	  \bibitem{congs} H. -S. Kim, H. Kim, J. Paek and S. Bahk, ``Load balancing under heavy traffic in rpl routing protocol for low power and lossy networks," \textit{IEEE Trans.  Mobile Comput.}, vol. 16, no. 4, pp. 964-979, Apr. 2017.
	   
	   \bibitem{C0} P. Fabian, A. Rachedi, C. Gueguen and S. Lohier, ``Fuzzy-based objective function for routing protocol in the internet of things," in \textit{Proc. IEEE GLOBECOM 2018}, Dec. 2018, pp. 1-6.
	   
	   \bibitem{C1} Chowdhury, A. Benslimane and C. Giri, ``Non-cooperative gaming for energy-efficient congestion control in 6Lowpan," \textit{IEEE Internet of Things J.}, vol. 7, no. 6, pp. 4777-4788, Jun. 2020.
	   	   
	   \bibitem{C2} S. Taghizadeh, H. Bobarshad and H. Elbiaze, ``CLRPL: Context-aware and load balancing rpl for iot networks under heavy and highly dynamic load," \textit{IEEE Access}, vol. 6, pp. 23277-23291, Apr. 2018.
	   
	   \bibitem{C3} S. L. Sampayo, J. Montavont and T. No\"{e}l, ``LoBaPS: Load balancing parent selection for rpl using wake-up radios," in \textit{Proc. IEEE ISCC 2019}, Jul. 2019, pp. 1-6.
	   	   
	   \bibitem{ML2} H. Yao, X. Yuan, P. Zhang, J. Wang, C. Jiang and M. Guizani, ``Machine learning aided load balance routing scheme considering queue utilization," \textit{IEEE Trans Veh. Technol.}, vol. 68, no. 8, pp.  7987-7999, Aug. 2019. 
	   
	   \bibitem{ML3} B.-S. Roh, M.-H. Han, J.-H. Ham and K-I. Kim, ``Q-LBR: Q-learning based load balancing routing for uav-assisted vanet," \textit{Sensors}, vol. 20, no. 19, pp. 1-18, Oct. 2020. 
	   
	   \bibitem{ML4} H. Yao, X. Yuan, P. Zhang, J. Wang, C. Jiang and M. Guizani, ``A machine learning approach of load balance routing to support next-generation wireless networks," in \textit{Proc. IEEE IWCMC 2019}, Jun. 2019, pp.1317-1322.
	    
	   \bibitem{ML5} A. Musaddiq,\textit{et al.}, ``Reinforcement learning-enabled cross-layer optimization for low-power and lossy networks under heterogeneous traffic patterns, " \textit{Sensors}, vol. 20, no. 15, pp. 1-26, Jul. 2020. 
	   
	   \bibitem{Q-learning} M. L. Littman, ``Reinforcement learning improves behavior from evaluative feedback," \textit{Nature}, vol. 521, pp. 445-451, May. 2015.
	   
	    
	   \bibitem{etx} JP. Vasseur \textit{et al.}, \textit{Routing metrics used for path calculation in low-power and lossy networks}, IETF RFC 6551, 2012.
	   
	   \bibitem{thunder} J. Hou, J. Rahul and Z. Luo, \textit{Optimization of parent-node selection in rpl-based networks}, IETF RFC 6921, Mar. 2017.
	   
	   \bibitem{std} J. Miranda \textit{et al.}, ``Path loss exponent analysis in Wireless Sensor Networks: Experimental evaluation," in \textit{Proc. IEEE INDIN 2013}, Jul. 2013, pp. 54-58.
	  
		
			\end{thebibliography}
\end{document}